\documentstyle[11pt,aaspp4,psfig]{article}

\newcommand{\chisq}{$\chi^2$}
\newcommand{\etal}{et al.\ }
\newcommand{\Msun}{$M_{\sun}$}
\newcommand{\Msunpyr}{\Msun yr$^{-1}$}

\begin{document}

\title{{\bf A dusty X-ray absorber in the Perseus Cluster ?}}

\author{K.A. Arnaud\altaffilmark{1} \&  R.F. Mushotzky}

\affil{Laboratory for High Energy Astrophysics, NASA/GSFC, Greenbelt MD 20771}
\altaffiltext{1}{also Astronomy Department, University of Maryland}

\slugcomment{Submitted to The Astrophysical Journal}

\lefthead{ARNAUD AND MUSHOTZKY}
\righthead{DUSTY X-RAY ABSORBER IN PERSEUS CLUSTER}

\begin{abstract}

We have analyzed 0.35--7.5 keV X-ray spectra of the center of the
Perseus cluster collected using the Broad Band X-Ray Telescope (BBXRT)
on the Astro-1 mission. These spectra are particularly useful for examining 
the nature of the X-ray absorber in cooling flows because of BBXRT's 
sensitivity between 0.35 and 1.0 keV. We confirm that there is X-ray
absorption above that expected from gas in our own galaxy. Further, 
the absorbing medium is deficient in helium. However, the energy 
of the K-edge of oxygen is consistent with neutral material (at the
redshift of the cluster) and is not consistent with any ionized state
of oxygen. It is not possible to completely ionize helium and have
oxygen neutral so the apparent helium deficiency cannot be due to
ionization. We propose that the X-ray absorption 
is due to dust grains that have condensed out of a medium in which 
helium remains ionized. This model satisfies all the observational
constraints but is difficult to understand theoretically.

\end{abstract}


\section{Introduction}

Clusters of galaxies are prodigious sources of X-ray emission,
produced by a $10^7-10^8$~K plasma confined by the gravitational
field of the cluster. In the inner tens of kpc of many clusters
the radiative cooling time of this gas is less than the age of
the cluster implying that the gas cannot be static and must
be cooling and flowing into the center (see eg Fabian 1994 and
references therein). This picture has been verified by both X-ray
imaging and spectroscopy with the implied cooling rates being, in some
cases, many hundreds of Solar masses per year.

This led to the central mystery of cooling flows : if so 
much mass is cooling through the X-ray band ($T\sim 10^7$K) what 
happens to it 
after that ? Solutions can be roughly classified into three 
camps : a) the X-ray data analysis is wrong, b) there is energy 
input which prevents the gas cooling out of the X-ray regime, 
and c) the gas cools down and condenses into something that is 
not observed (eg low mass stars). Solution a) has been shown
to be incorrect by the consistent results obtained using a variety
of different detectors and analysis methods. Solution b) is
much argued over and seems to us to require an unlikely amount
of fine tuning. Solution c) is unsatisfactory in that it posits
placing the mass in something that cannot be seen. There is some
evidence for star formation in the galaxies at the centers of
cooling flows but with a conventional initial mass function this
can mop up only 1--10\%
of the available mass (e.g. McNamara 1997).

Solution c) received a large boost following the observation
by White \etal (1991) that cooling flows showed X-ray absorption
in excess of that from our own Galaxy. The column densities observed
were consistent with the accumulation of cooling flow matter over
the lifetime of the cluster. These excess absorptions were confirmed
by analysis of spectra obtained using ROSAT (Allen \etal 1993) and
ASCA (Fabian \etal 1994). However,
this cold, absorbing material is not seen
in HI or CO (e.g. O'Dea \& Baum 1997). Again, we obtain a 
result using X-ray data
but cannot find the corollaries in other wavebands. A further problem
is that the excess absorption does not always show up in the
analysis of spectra obtained using the ROSAT PSPC (e.g. Sarazin
1997). Since the ROSAT PSPC has a lower energy cut-off than ASCA it 
should provide a more sensitive measurement of absorption. 

In this paper we report investigations on the nature of the X-ray
absorber in the Perseus cluster, a much-studied, massive cooling flow
centered on the galaxy NGC 1275.
We use data from the Broad Band X-Ray Telescope (BBXRT) that flew
on the Astro-1 Space Shuttle mission December 2--10, 1990. BBXRT used 
telescopes similar
to those on ASCA but with segmented Si(Li) solid-state
detectors instead of CCDs. The BBXRT Si(Li) detectors have slightly
lower spectral
resolution than the ASCA CCDs but they do not have the instrumental
oxygen absorption which reduces the CCDs' efficiency immediately
above 0.5 keV. Since the primary absorption feature in the energy
range covered by these detectors is the O K edge, BBXRT data provide a 
more sensitive measurement of the redshift and detailed shape of the 
X-ray absorption than is possible with ASCA.

\section{Instrument and Observation Description}

BBXRT (Serlemitsos \etal 1991,
Petre \etal 1991, Weaver \etal 1995) was flown on the Space Shuttle 
Columbia (STS-35) as part of the nine day Astro-1 mission 
(Blair \& Gull 1990). BBXRT 
consisted of two grazing-incidence conical foil telescopes with focal 
lengths of 3.8m. These telescopes had a spatial resolution whose
integral power can be characterized by two error functions : 65\% 
of the reflected photons are distributed with an half-power diameter 
(HPD) of 1.3 arcminutes and the rest with an HPD of 5.8 arcminutes. 
The system was sensitive to X-rays with energies up to 12 keV, an upper
limit determined by the focal length, although the effective area 
dropped off rapidly as this energy was approached since there were few 
foils available to focus X-rays requiring such small grazing angles.
The mirrors were prototypes for those used on ASCA and have similar
properties.

The focal plane instrumentation consisted of two segmented
Si(Li) solid-state detectors. Each detector was divided into five
independent pixels, a central 4 arcmin diameter pixel and four outer 
quadrant pixels covering a total diameter of 17 arcmin with 1
arcmin gaps between individual pixels. We refer to the central
detector segments as A0 and B0 with the A system being that for the 
telescope whose axis is taken as the nominal pointing direction. The
telescopes were misaligned by $\sim 1'$.
The
outer quadrant pixels for the A system are labelled A1 -- A4 in a 
anticlockwise direction on the sky with the roll angle defined by the angle
between due East and the A1/A2 dividing line. The B outer pixels are
rotated by 180 degrees with respect to the A system so B1 corresponds
to A3, B2 to A4 etc. The central pixels had an energy 
resolution of $\sim 165$eV (FWHM) at 6 keV and the outer pixels a
resolution
of $\sim 190$eV. Each detector returned spectra in 512 channels, the
first 256 being approximately 16 eV wide and the rest 32 eV wide.
Background rejection was performed using a combination
of criteria, the most important of which were pixel-pixel coincidence 
and the triggering of a wrap-around plastic scintillator.

BBXRT pointed in the direction of the Perseus cluster on four
occasions during the Astro-1 mission (there were several more
very brief observations). The observation analyzed here was the
longest and highest signal-to-noise. It occurred on 
day 6 of the mission. Star-tracker data shows that the pointing 
position was $\alpha=49.12$, $\delta=41.38$, approximately 3 arcminutes
from NGC 1275. This placed the center of the cluster
on a gap between pixels. The roll angle was 120 degrees so the 
cooling flow region was distributed between pixels A0 and A3 on one 
telescope and B0 and B1 on the other. The exposure time was 2542.5 
seconds. The observation on day 3 was also pointed at the center 
of the cluster however its exposure time of 330 seconds is too short 
to provide useful data.

We used the source spectra, background spectra, and response matrices
from the HEASARC archive. These were generated using the standard
procedures given in Weaver \etal (1995). We note, in particular, that
this observation occurred during orbit night so there is no danger
of contamination due to an atmospheric oxygen line.

In our analysis we used channels starting at 0.35 keV for A0 and B0, 
0.44 keV for A3 and 0.60 keV for B1. The outer pixels were less
reliable at very low energies, and those on the B telescope
particularly so (Weaver \etal 1995). We cut off all the spectra at 
7.5 keV because of the sharp decrease in telescope effective area
above this energy. We grouped the spectra so that each bin contained 
at least 20 counts (although this only affected the extreme high 
energy end). We subtracted background although the effect is negligible.

In our analysis we fitted models to all four spectra simultaneously
but left their relative normalizations as free parameters.

\section{Results of fits to the data}

We started by fitting a simple series of models to the data in
order to test if we were in agreement with the results from
the ASCA SIS published by Fabian \etal (1994). We
used XSPEC v10 (Arnaud 1996) to fit the following series of models to the
data. First, a single temperature Raymond-Smith plasma 
(RS: Raymond \& Smith 1977) with photoelectric absorption from the
Galactic column ($1.37\times 10^{21}$ cm$^{-2}$: Stark \etal 1992). 
The data (for pixel A0 only) and this model are shown in Figure 1.
The deficit of model flux between 1 and 2 keV indicates that an
additional lower temperature component is required. However, the
model lies above the data between 0.5 and 1 keV so extra absorption
is also necessary. We
added a second component with its own absorption, assumed to
be at the redshift of the cluster ($z=0.0183$). This second
component was, successively, another single temperature RS plasma,
a cooling flow model consisting of emission from a range of
temperatures (Mushotzky \& Szymkowiak 1988), and a cooling flow
model using the Mewe-Kaastra-Liedahl plasma model (MK: Mewe, 
Gronenschild \& van den Oord 1985; Mewe, Lemen \& van den Oord 1986; 
Kaastra 1992, Liedahl, Osterheld \& Goldstein 1995).
The results are shown in table 1. None of the models provides a
good fit to the BBXRT data (as was true with the ASCA data). 
The fit parameters are in broad agreement between the two sets of 
detectors. The amount of extra absorption required is the largest difference 
between the BBXRT and ASCA results. This parameter is determined 
by precisely the energy range where the BBXRT and ASCA sensitivities 
differ the most.

\vspace{0.1in}
\begin{minipage}{3.5in}
\centerline{\psfig{figure=f1.eps,width=3.4in,height=3.4in,angle=270}}
\end{minipage}
\begin{minipage}{2.5in}
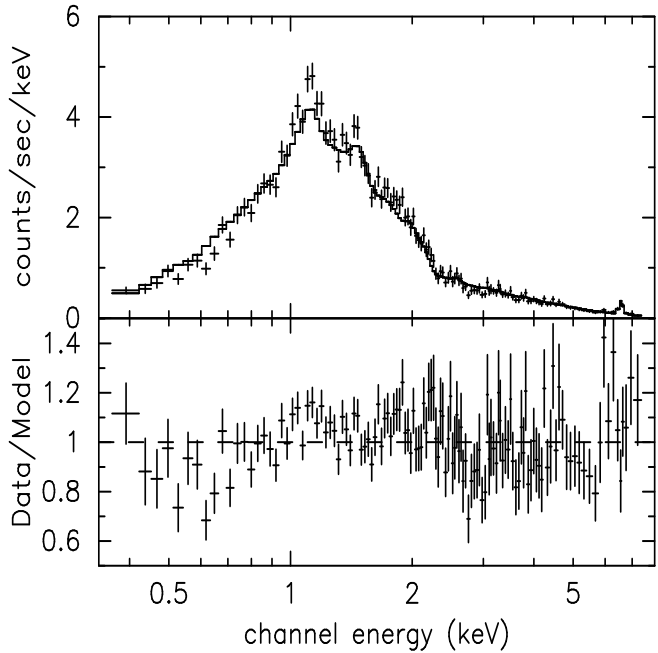
Figure 1. The upper panel shows the A0 data (crosses) and model
(stepped line) for the single temperature fit. The lower panel shows
the data divided by the model. The vertical bars on the data are one 
sigma errors.
\end{minipage}
\vspace{0.1in}

The strongest absorption feature in the energy range covered by
the BBXRT data is due to the oxygen K edge. So, we replaced the
redshifted absorption with an edge and found its best-fit energy
and confidence range. The result for the model using the RS cooling 
flow is shown in Figure 2. We have checked the channel-energy
relation by fitting an oxygen fluourescence line to BBXRT bright
Earth data. We find an energy of $528\pm 1$eV, which is 3eV above
the theoretical value. The bands marked z=0 and z=0.0183 on Figure 2
are the edge energies shifted by 3eV to correct for this slight
miscalibration
\footnote{There is some dispute about the energy of the oxygen K
edge. The value used in the photoelectric absorption routines commonly
used in X-ray astronomy (e.g. Balucinska-Church \& McCammon 1994) is
532 eV. However, Schattenburg \& Canizares (1986) place a lower
limit on the edge energy of $\sim 536$ eV using FPCS observations 
of the Crab nebula. These authors note that a possible explanation
for the discrepancy is that the 532 eV value given in the literature
is the energy required to promote a K-shell electron to the Fermi
level, rather than to the continuum. This could produce a 
$\sim 5$ eV shift. Indeed, Verner \etal (1996) place the edge at 537 eV.
However, Gould \& Jung (1991) argue that
the current experimental measurements are not relevant to the
astrophysical context and their theoretical calculation gives an
edge energy of 546 eV. The latest results from the Opacity Project
give an edge energy of 549 eV (Pradhan \& Zhang, priv.comm.)
We calibrated the gain of BBXRT assuming a fluourescent line energy of
525 eV. This corresponds to an edge at 537 eV. If the edge is at a higher
energy then so is the fluourescence line and our gain calibration is
changed. Thus any result on the redshift of the edge is independent of
the actual correct edge energy.
We did modify the Balucinska-Church
\& McCammon (1994) subroutine to use 537 eV. We also tested our
results using the cross-sections of Verner \etal (1996) and found no
significant changes.}.

\vspace{0.1in}
\begin{minipage}{3.5in}
\centerline{\psfig{figure=f2.eps,width=3.4in,height=3.4in,angle=270}}
\end{minipage}
\begin{minipage}{2.5in}
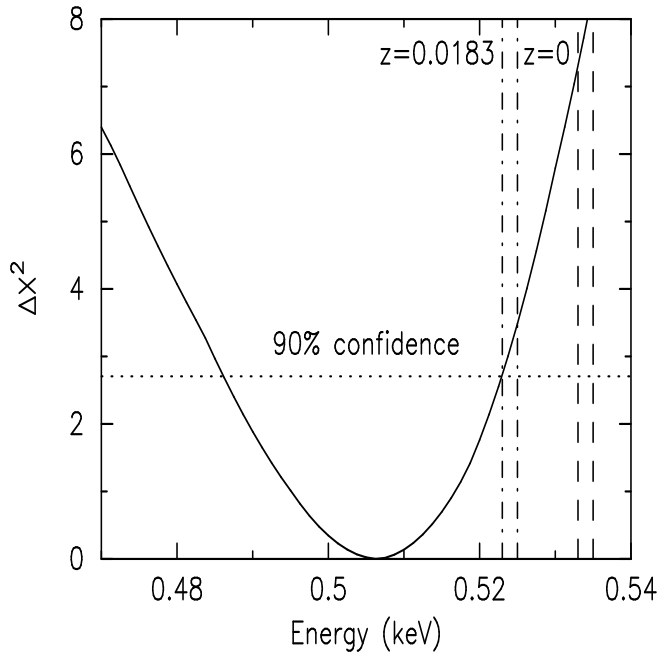
Figure 2. The confidence region for the edge energy from fitting
an RS cooling flow model and edge to the BBXRT data. The horizontal dotted
line is at $\Delta\chi^2=2.71$, corresponding to 90\% confidence on one
interesting parameter. The vertical bands are edge energies (with calibration
uncertainties) for an absorber at the redshift of the cluster and at zero
redshift.
\end{minipage}
\vspace{0.1in}

The widths of the bands reflect the uncertainty in the gain calibration.
Formally, we can rule out at $>$90\% 
confidence the oxygen edge being due to material in the Galaxy. However,
there are 5-10\% 
uncertainties in the effective area calibration in this energy range 
(Weaver \etal 1995) which must affect the measurement of the edge 
energy. So, a conservative statement would be that the edge energy is 
better fit by an absorber in the rest frame of the cluster but we
cannot rule out a zero redshift absorber. The K edge of singly-ionized
oxygen is $\sim 20$ eV above that for neutral material. An edge at
this energy is ruled out by the BBXRT data so the oxygen in the
absorbing material must be neutral.

The model with a single oxygen edge is a significantly better fit
to the data than that with standard photoelectric absorption. For the 
RS cooling flow model the former gives \chisq=1185 and the latter 
\chisq=1211 . We will refer to these two models as the ``edge'' and
``absorption'' models respectively. The difference between the fits
is illustrated in Figure 3, which shows the ratios of data to best-fit
model for the two cases. The ``absorption'' model leads to a
systematic slope to the residuals between 0.4 and 0.6 keV while the
``edge'' model gives flat residuals in this energy range.

\vspace{0.1in}
\begin{minipage}{3.5in}
\centerline{\psfig{figure=f3.eps,width=3.4in,height=3.4in,angle=270}}
\end{minipage}
\begin{minipage}{2.5in}
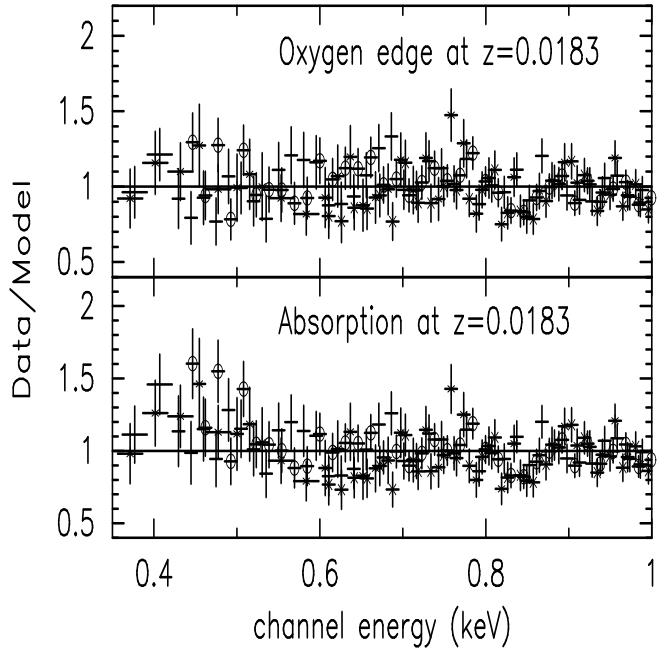
Figure 3. The upper panel shows the ratio of data to best-fit model
for a single edge modifying the cooling flow while the lower panel is for
a standard photoelectric absorption model modifying the cooling flow. The
four detector pixels are indicated by different symbols.
\end{minipage}
\vspace{0.1in}

We have checked that the slope in the ``absorption'' model residuals
cannot be due to systematic problems in the instrument calibration by 
looking at the BBXRT observations of the
Crab nebula (see also Weaver \etal 1995). We have compared the ratios
of central pixel (A0 and B0) data to best-fit model for the Perseus
and Crab nebula observations. Figure 4 shows the ratio of data to
absorbed power-law model for the Crab (circle symbols) and the ratio
of data to ``absorption'' model for the Perseus cluster. The Perseus
cluster residuals differ systematically from those for the Crab nebula
so the ``absorption'' model residuals cannot be due to a calibration
problem.

In our attempts to find a good model for the data including standard
photoelectric absorption we have tried changing from a RS-based cooling 
flow to one based on MK. We have also tried varying the assumed minimum 
temperature of the flow and the slope of the differential emission 
measure function (Mushotzky \& Szymkowiak 1988). None of these changes 
improve the fit below 0.7 keV. The reason why none of these changes work
can be seen in Figure 5. This figure shows the ``absorption'' model 
used to give the residuals in the lower panel of 
Figure 3. The lower line is the cooling flow component modified by its 
absorption, the middle line is the single temperature RS model, and the 
top line is the sum of the two. The cooling flow component does not
contribute significantly to the model below about 0.7 keV. Hence,
modifying this component cannot improve the fit in the 0.4--0.6 keV
range. 

\vspace{0.1in}
\begin{minipage}{3.5in}
\centerline{\psfig{figure=f4.eps,width=3.4in,height=3.4in,angle=270}}
\end{minipage}
\begin{minipage}{2.5in}
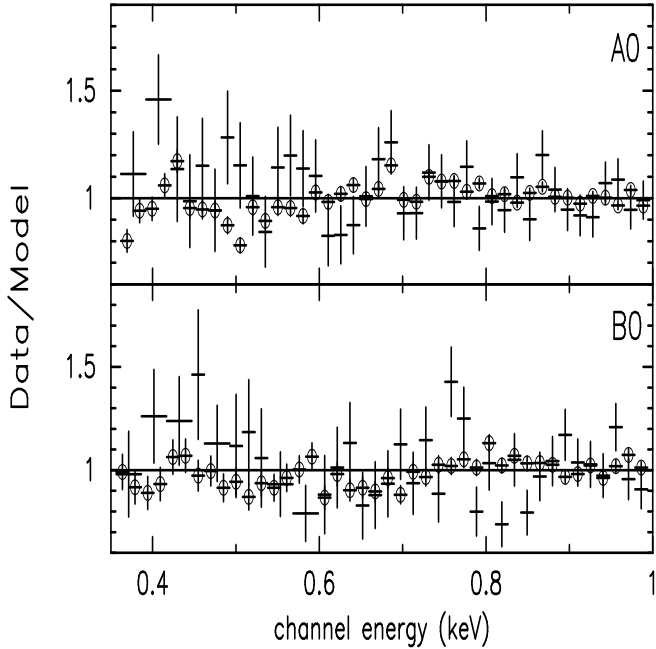
Figure 4. The points indicated by circle symbols and with smaller error
bars are the ratio of the data to best-fit power-law model for the
Crab nebula. The other points are the ratio of the Perseus cluster
data to the cooling flow modified by the standard photoelectric
absorption model (ie the lower panel in Figure 3). The upper panel is
for detector pixel A0 and the lower for B0. Note that the Perseus
cluster points are systematically high relative to the Crab points for
energies between 0.4 and 0.6 keV.
\end{minipage}
\vspace{0.1in}

We also note that although the 2-temperature 
model with absorption gives a lower total \chisq\ than the cooling flow with 
an edge the residuals below 0.7 keV still show the systematic problem. 
Indeed, a 2-temperature model with an edge gives \chisq=1166, the lowest of 
any model we tried.

The ``absorption'' model used above assumes that the cooling
flow region is in the center of the cluster, outside that there is a 
skin of cold absorbing material, and outside that the ambient medium,
represented by the single temperature RS model. This is physically
implausible. There are a large number of ways that the absorbing
medium can be distributed; we have looked at two representative
examples. Firstly, we suppose that some fraction of the cooling flow
is outside the absorbing material and the rest is inside. This can be
described by a partial covering fraction model. Attempting to fit this
model to the data gave a best fit with unit covering
fraction. Alternatively, we supposed that the absorbing clouds are
distributed uniformly throughout the cooling flow region. Allen \& 
Fabian (1997) called this the multilayer absorber model. It
has a transmission given by $(1 - e^{-\sigma(E)N_H})/(\sigma(E)N_H)$,
where $\sigma(E)$ is the absorption cross-section and $N_H$ is the
total column through the cooling flow region. We fit this model
to the data and found that it did even worse than the simple
absorption model (\chisq=1229 vs. \chisq=1211) and did not solve
the problem at low energies.

Figure 6 illustrates why these modifications to the ``absorption'' model
do not work. The solid curve is the ``absorption'' model and the
dashed curve is ``edge'' model. Almost the only difference is that
the ``edge'' model gives more flux in the range 
0.4--0.5~keV. The modified absorption models do allow more flux through 
in this energy range but at the expense of flattening the spectrum in the 
range 0.5--1.0~keV, which the data do not allow.

\vspace{0.1in}
\begin{minipage}{3.5in}
\centerline{\psfig{figure=f5.eps,width=3.4in,height=3.4in,angle=270}}
\end{minipage}
\begin{minipage}{2.5in}
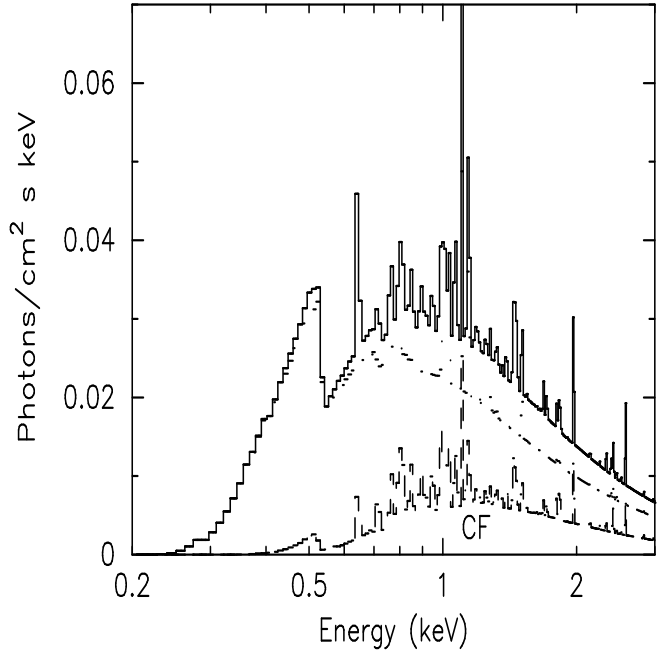
Figure 5. The theoretical model components for the photoelectric 
absorption case shown in the lower panel of figure 3. The lower 
(dashed - labelled CF) line is the cooling flow component modified by 
its absorption, the middle (dashed) line is the single temperature RS 
model, and the top (full) line is the sum of the two.
\end{minipage}
\vspace{0.1in}

The only option left is to change the relative abundances of elements
in the absorber. To get an idea of which abundances mattered the most
we first set the abundances of all elements other than O to zero. Then
we set each
element in turn to Solar abundance (Anders \& Grevesse 1989)
relative to O and fit for the
other free parameters. The elements that produced the greatest
increase in \chisq\ were Ne, He, H, C, Mg, Fe with $\Delta$\chisq\ 
of +15, +14, +6, +6, +5, +5 respectively. The Ne K-edge is at 
$\sim 0.87$ keV, in the energy range where the cooling flow makes
a significant contribution. In view of this and the systematic
differences between the data and the model in this energy range
visible in figure 3, we prefer not to draw strong conclusions from the
lack of Ne absorption. However, there are no systematic problems below
the O K-edge, where absorption is dominated by He. Consequently, we 
consider the He result to be firm. Thus the data imply a large
deficiency in He abundance with respect to O. We then
assumed that H, He, and Ne have their standard relative abundances and
then fit for their abundance relative to oxygen. We found that the 
combination of H, He, and Ne must be underabundant relative to O by a
factor of 10 with respect to Solar abundances. We performed a similar
test for C and found that it must be underabundant relative to O by a
factor of 3.

\vspace{0.1in}
\begin{minipage}{3.5in}
\centerline{\psfig{figure=f6.eps,width=3.4in,height=3.4in,angle=270}}
\end{minipage}
\begin{minipage}{2.5in}
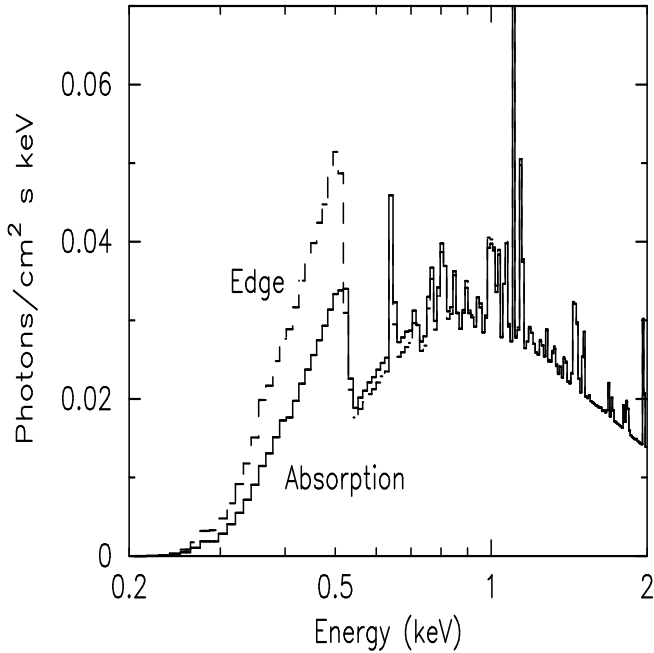
Figure 6. The two theoretical models used in figure 3 plotted on the same 
graph. The solid line is for standard photoelectric absorption and the
dashed line is for the single edge.
\end{minipage}
\vspace{0.1in}

\section{Discussion}

The simplest way to produce O absorption without He
absorption is to suppose that the He is completely ionized. However,
the O K-edge seen in the BBXRT spectrum is from neutral material. We can
rule out even singly-ionized oxygen, whose K-edge is $\sim 20$~eV
higher than that of neutral oxygen. It is not possible to have both
neutral O and completely ionized He either for collisional
(e.g. Arnaud \& Rothenflug 1985) or photoionized (e.g. Kallman \& McCray 1982)
equilibrium plasmas. So, we can rule out ionization as a mechanism for
the deficit in absorption due to He.

The argument in the previous paragraph holds only if the He and O are
mixed together in the gaseous state. One way to combine neutral O
and ionized He is to suppose that the O is in grains that have condensed
out of the gas. This explanation is bolstered by the result on Ne absorption.
He and Ne are noble gases and are not depleted onto grains. 

It would be overinterpreting data of this signal-to-noise and
resolution to attempt to deduce abundance ratios in the grains.
We merely comment that we do not expect the absorber to have Solar 
abundance ratios. Mushotzky \etal (1996) show that the relative 
abundances in the intracluster medium of O, Ne, Mg, Si, S, and Fe
are non-Solar and are consistent with production in Type II
supernovae. There are no observations giving the abundance of C in 
intracluster gas. Since C is not a Type II supernova product we might
expect it to be underabundant with respect to O.

An He deficiency provides a possible explanation for 
discrepancies found between measurements using the ROSAT PSPC
and measurements using other detectors. The determination of
an absorption column using the PSPC is dominated by data in the
energy range 0.2--0.5 keV (see for instance figure 1 in Sarazin 1997).
The absorption in this energy range is mainly due to He.
Since the cooling flow absorber in the Perseus cluster is apparently 
deficient in this element the
PSPC will be insensitive to the extra absorption and will tend to measure
only the Galactic absorption column. The significant extra absorption
in the Perseus cluster measured by Allen \& Fabian (1997) was based 
on PSPC 0.41--2.00 keV data. In this energy range using the PSPC
oxygen is the dominant absorber so extra absorption is detected. 
Allen \& Fabian (1997) do
consider energies down to 0.2 keV for the spectra from A1795 and
A2199. Using these data they find smaller excess columns than if they
used spectra starting at 0.4 keV. We interpret this as evidence for
a He deficiency in the absorber in these two clusters. Allen \& Fabian
(1997) did not use the low energy data in general because the ROSAT
PSPC spatial resolution is worse than at higher energies and they were 
interested 
in the spatial distribution of the absorber. We have analyzed the
entire ROSAT PSPC spectrum of the whole cooling flow region in Perseus
(which is large enough that we don't have to worry about the energy-dependent
spatial resolution). For a wide variety of emission models we find that the
absorption is consistent with the Galactic column. Although there are
problems with the calibration of the ROSAT PSPC (Snowden priv. comm.)
these do not affect the measurement of the absorption column. This provides
confirmatory evidence for absorber He deficiency found using the BBXRT data. 
The poor spectral resolution
and large instrumental carbon edge make the sort of analysis performed
in this paper extremely difficult using the PSPC. 

The presence of dust around NGC 1275 has been deduced from dark regions in
optical images
(McNamara, O'Connell \& Sarazin 1996), the ratios of emission lines in 
the optical filaments (Hu 1992; Donahue \& Voit 1993), and far-IR 
emission (Lester \etal 1995). The observed X-ray absorption column
and cooling flow luminosity require the grains to absorb about
$5\times 10^{43}$ erg/s (we use H$_0 = 50$ km/s/Mpc throughout). 
The exact absorbed luminosity depends on
the distribution of absorption and emission in the BBXRT
field-of-view. The 40--120\micron\ luminosity quoted by Lester \etal
(1995) is double this absorbed X-ray luminosity. So the absorbed X-ray
energy could be reradiated in the far-IR.

The O column deduced from the X-ray absorption is $\sim 2\times 
10^{18}$~atoms/cm$^2$. If the absorber is associated with the cooling
flow and is distributed across the cooling region this gives a mass in O of
% 2e18*2e47*16*1.7e-24/2e33   
$\sim 4\times 10^9 f$ \Msun, where $f$ is the covering fraction of
the cooling flow and absorption. For $f \sim 1$ this is a mass in
grains two orders of 
magnitude larger than that deduced from the far-IR emission (Jura \etal 1987; 
Lester \etal 1995). However, the far-IR emission is not distributed
across the cooling flow but appears to correlate with the emission
line filaments (Lester \etal 1995). We are led to one of two
conclusions: a) the cooling flow emission and its absorbing 
dust cover $\sim 1$\% 
of the cooling region and correlate with the emission 
line filaments, or b) the cooling flow emission and absorbing dust are
distributed across the cooling region but outside the emission line
filaments the dust is too cold to radiate significantly at 
100\micron\ (ie T$\lesssim 10$K).
Option a) is ruled out by the high spatial resolution X-ray observations
which do not show any correlation between X-ray emission and emission
line filaments (B\"ohringer \etal 1993; McNamara, O'Connell \& Sarazin
1996).

We emphasise that it is the lack of correlation between the X-ray
emission and the optical filaments that it is critical. If the X-ray emission
and absorption from the cooling flow had been associated with the
filaments then we could have built a coherent picture involving
absorption due to the warm dust and molecules which are seen (Lester
\etal 1995; Lazareff \etal 1989; Mirabel \etal 1989). We can reduce
the mass in dust required by supposing that the cooling flow and its
absorber are distributed throughout the cooling region but
are inhomogeneous on a scale smaller than the resolution of the ROSAT
HRI. There is no observational data on the covering fraction at these
spatial resolutions. There is evidence in the Centaurus clusters for
a volume filling factor less than one. Fukazawa \etal (1994) showed 
that in the cooling region the lower temperature component occupied a 
volume one twentieth of that of the hotter ambient medium. The
relation between the volume filling factor and the covering fraction
will depend on the details of the cooling flow.

The identification of the X-ray absorber with cold dust explains all
the observations but poses severe theoretical problems. Dust should
be destroyed by sputtering in the central regions of the cluster 
(Dwek \& Arendt 1992).
Some of the dust could be in the outer parts of the cluster (e.g. Hu
1992) however Allen \& Fabian (1997) show that the excess absorption is 
concentrated in the core of the cluster, within the cooling region.
So, we require that cold ($\lesssim 10$K) dust co-exist in the cooling
region with gas of densities $> 10^{-2} $cm$^{-2}$ and temperatures of
$\sim 10^7$~K. Further, if the dust is formed from the cooling flow
then the gas needs to cool below the condensation temperature ($\sim
2000$~K) but be ionized such that He is completely stripped. The
optical spectrum of the stars in NGC 1275 is not reddened above that 
expected due to dust in our own galaxy. This requires that the dust
grains be large enough to be opaque. Since the timescale for dust 
destruction by sputtering is proportional to grain size, the smaller 
grains, which provide the optical reddening, will not survive.

Dust appears to form easily whenever the gas temperature drops below 
the condensation temperature for grains (e.g. SN1987A: Colgan \etal
1994) and in AGN grains co-exist with ionized gas (e.g. Brandt, 
Fabian \& Pounds 1996). Fabian, Johnstone \& Daines (1994) argued 
that cold, dusty clouds can form in cooling flows. However, these 
clouds will also contain cold He so will absorb X-rays below the 
neutral O K-edge. Voit \& Donahue (1995) pointed out that CO would
be evaporated off grains by transient X-ray heating from the cluster
gas. In this case most of the oxygen would not be on grains but
in molecular gas whose luminosity would exceed that observed in the
CO rotational lines. Also, heating of the grains by collisions with
hot electrons from the cluster gas is capable of supplying more than
enough energy to surpass the far-IR luminosity observed (Lester \etal
1995). We have no solution to these problems.

We note in passing that Czerny \etal (1995) considered X-ray
absorption by dust grains in AGN. They showed that if dust rather
than gas is the absorbing medium then more flux will be detected
around 0.4 keV and this is may explain some observations of
soft excesses above a power-law spectrum absorbed by gas.

\section{Conclusions}

We have shown that the BBXRT spectra of the Perseus Cluster are
consistent with a cooling flow absorber at the redshift of the
cluster. The data are marginally inconsistent with the absorber
being local to our galaxy. More important, we have demonstrated
that the absorber is deficient in helium (and, with less confidence,
neon). The only reasonable explanation of this is that the
absorber is comprised of dust grains formed out of gas in which
helium is completely ionized. The absorption model we 
have fit assumes that the dust grains are a screen in front of the 
cooling flow. This is oversimplistic and it is likely that they are
distributed throughout the cooling flow. The present data are not
good enough to explore this issue.

A model in which cold grains are distributed throughout the cooling
region and are heated only where there are emission line filaments
satisfies all the observational constraints. However, we do not
understand how to keep the oxygen on the grains, how to prevent the 
grains being destroyed, or how to keep the grains at very low effective
temperatures.

BBXRT also observed the cooling flow clusters Abell 262 and Abell 496.
These data are not as high quality as those for Perseus and all we can
conclude for these clusters is that there is an O K edge present whose
depth exceeds that expected from the Galactic column (Mackenzie,
Schlegel \& Mushotzky 1996).

Our result could be confirmed using the 
the backside-illuminated ACIS CCDs on AXAF. In the future, high resolution
spectroscopy will provide detailed information about abundances,
ionization states, and grain composition from the energies, depths,
and detailed shapes of absorption edges (see e.g. Woo, Forrey \& Cho 1997).

We thank Ari Laor for reminding us that we had never published this
data. We thank Hagai Netzer and Tim Kallman for helpful discussions.
This research has made use of data obtained through the High Energy 
Astrophysics Science Archive Research Center Online Service, provided 
by the NASA/Goddard Space Flight Center.

\newpage

\def\res#1#2#3{$#1^{+#2}_{-#3}$}

\begin{deluxetable}{llll}
\tablewidth{0pt}
\tablecaption{Results of spectral fits}
\tablehead{Model and Parameters & Fit values & (F94)\tablenotemark{a} & \chisq/bins}
\startdata
Single temperature  &   &  &    1367/1043 \nl
\quad\quad\quad Temperature & \res{4.22}{0.07}{0.07} keV & (4.15) & \nl
\quad\quad\quad Abundance   & \res{0.41}{0.03}{0.03} of Solar& (0.48) &
\nl
Two temperature with absorption & & & 1179/1043 \nl
\quad\quad\quad Higher Temperature & \res{5.25}{0.57}{0.33} keV&
(5.62) & \nl
\quad\quad\quad Lower Temperature  & \res{1.57}{0.25}{0.21} keV&
(2.05) & \nl
\quad\quad\quad Extra absorption  & \res{0.22}{0.12}{0.08} $\times
10^{22}$cm$^{-2}$ & (0.11) & \nl
\quad\quad\quad Abundance  & \res{0.37}{0.04}{0.03} of Solar& (0.39) &
\nl
Cooling flow (RS) with absorption & & & 1211/1043 \nl
\quad\quad\quad Temperature & \res{4.29}{0.17}{0.15} keV& (4.21) & \nl
\quad\quad\quad Extra absorption  & \res{0.29}{0.04}{0.03} $\times
10^{22}$cm$^{-2}$ & (0.41) & \nl
\quad\quad\quad Abundance  & \res{0.40}{0.03}{0.03} of Solar& (0.46) &
\nl
\quad\quad\quad Accretion rate & \res{180}{26}{26} \Msunpyr & (138) & \nl
Cooling flow (MK) with absorption & & & 1211/1043 \nl
\quad\quad\quad Temperature & \res{4.20}{0.15}{0.14} keV& (4.20) & \nl
\quad\quad\quad Extra absorption  & \res{0.35}{0.06}{0.04} $\times
10^{22}$cm$^{-2}$ & (0.30) & \nl
\quad\quad\quad Abundance  & \res{0.39}{0.03}{0.04} of Solar& (0.43) &
\nl
\quad\quad\quad Accretion rate & \res{154}{23}{23} \Msunpyr & (122) & \nl
 & & & \nl
Two temperature with O K-edge & & & 1166/1043 \nl
\quad\quad\quad Higher Temperature & \res{5.85}{0.48}{0.68} keV& &\nl
\quad\quad\quad Lower Temperature  & \res{2.02}{0.37}{0.44} keV& &\nl
\quad\quad\quad Optical depth & \res{1.88}{1.86}{0.56} & &\nl
\quad\quad\quad Abundance  & \res{0.38}{0.03}{0.04} of Solar& &\nl
Cooling flow (RS) with O K-edge & & & 1185/1043 \nl
\quad\quad\quad Temperature & \res{4.49}{0.19}{0.16} keV& &\nl
\quad\quad\quad Optical depth & \res{4.30}{0.55}{0.46} & &\nl
\quad\quad\quad Abundance  & \res{0.40}{0.03}{0.04} of Solar& &\nl
\quad\quad\quad Accretion rate & \res{180}{23}{24} \Msunpyr & &\nl
\enddata
\tablenotetext{a}{Analogous results from the ASCA SIS analysis by
Fabian \etal (1994). The regions of the cluster from which spectra
are accumulated are different for the ASCA SIS and BBXRT so these
results are not precisely comparable. In particular, the measured mass
accretion rates should differ.}
\end{deluxetable}

\end{document}